\documentclass[a4paper,11pt]{article}
\pdfoutput=1 
\usepackage[utf8]{inputenc}
\usepackage{jheppub} 
\usepackage{amsmath}
\usepackage{bbold}
\usepackage{xcolor}
\usepackage{subfig}
\usepackage{graphicx}
\usepackage{hyperref}
\usepackage{mathtools}
 \usepackage{mdframed}

\newcommand{\RN}[1]{%
  \textup{\uppercase\expandafter{\romannumeral#1}}%
}

\newcommand{\bh}{\bar{h}}

\def\hb{\frac{\beta}{2}}

\def\a{\alpha}
\def\b{\beta}
\def\c{\chi}
\def\d{\delta}

\def\i{\iota}
\def\j{\psi}
\def\k{\kappa}
\def\l{\lambda}

\def\o{\omega}

\def\r{\rho}

\def\t{\tau}

\def\D{\Delta}

\def\L{\Lambda}
\def\O{\mathcal{O}}
\def\P{\Pi}

\def\S{\Sigma}

\def\bz{\bar{z}}

\newcommand{\zb}{\bar{z}}

\newcommand{\hypf}{{}_2F_{1}}
\newcommand{\gii}{\texorpdfstring{${\cal G}(z,\bz)\big|_{\textrm{II}}$}{TEXT}}
\newcommand{\gi}{\texorpdfstring{${\cal G}(z,\bz)\big|_{\textrm{I}}$}{TEXT}}

\title{Correlators of Four Light-Ray Operators in CCFT}
\author[a]{Shounak De,}
\author[a,b]{Yangrui Hu,}
\author[a]{Akshay Yelleshpur Srikant,}
\author[a]{Anastasia Volovich}

\affiliation[a]{Department of Physics,
	Brown University,
	Providence, RI 02912, USA}
\affiliation[b]{Brown Theoretical Physics Center,
	Box S, 340 Brook Street, Barus Hall,
	Providence, RI 02912, USA}

\emailAdd{shounak\_de@brown.edu}
\emailAdd{yangrui\_hu@brown.edu}
\emailAdd{akshay\_yelleshpur\_srikant@brown.edu}
\emailAdd{anastasia\_volovich@brown.edu}

\abstract{Building on results in arXiv:2203.04255, we compute the correlator of four gluon light-ray operators in celestial CFT. We find that it is described by Fox H-functions and generalized I-functions of multiple variables. We also analyze light-ray correlators for a scalar amplitude involving the exchange of a massive scalar.
}

\begin{document}
	\maketitle
	\section{Introduction}

Light-ray operators appear in Lorentzian CFTs~\cite{Kravchuk:2018htv, Korchemsky:2021htm, Kologlu:2019mfz} in general, and in celestial CFT (CCFT) in particular \cite{Himwich:2021dau, Atanasov:2021cje}. Their role in CCFT is tied fundamentally to the problem of finding a suitable basis of operators.
%The reason is that their correlation functions are light-ray transforms of ordinary CCFT correlators.
It was shown in \cite{Sharma:2021gcz} that certain three-point correlators involving light-ray operators take the form of standard three-point CFT correlators. This is in contrast to the  distributional nature of CCFT three-point functions \cite{Pasterski:2017ylz}. These operators were further studied and their contribution to the OPE of CCFT primaries was identified in \cite{Guevara2021tvr}. Four-point correlators involving one and two gluon light-ray operators were computed in \cite{Hu:2022syq, Banerjee:2022hgc}. They are non-distributional and correctly reproduce the contribution of light-ray operators to the OPE. This contribution is invisible from the collinear limits of the four-point celestial amplitude. 

In this paper, we build on our results from \cite{Hu:2022syq} and compute correlators of four gluon light-ray operators. We find that they can be expressed in terms of Fox H-functions \cite{Hfunc1,Hfunc2,Hfunc3,Hfunc4,multiHfunc} and generalized I-functions \cite{Ifunc1,Ifunc2}. We also analyze scalar amplitudes involving the exchange of a massive scalar. 

The paper is organized as follows. In Section \ref{sec:four-light-gluon}, we compute the four light-ray gluon correlator. We flesh out the computation of light-ray correlators involving the exchange of a massive scalar in Section \ref{sec:scalar-light-transf}. 
The paper ends with a number of appendices which contain technical details of some computations as well as a variety of other results. Appendices \ref{appen:integral-gluon} and \ref{appen:integral-scalar} are devoted to providing explicit expressions for various integrals used throughout the paper. Appendix \ref{appen:I-func} defines the relevant special functions. Appendix \ref{appen:g1} describes the technical details involved in the evaluation of the four gluon light transform. Appendix \ref{appen:scalar-OPEs} contains the derivation of various OPEs from the collinear limits of the scalar light-ray correlator. Finally, Appendix \ref{appen:CBD-scalar} describes the conformal block decomposition of the scalar light-ray correlator.\\

\noindent {\bf Note}: During the preparation of this draft, we were aware of the upcoming work~\cite{unpublished}, which also computes four light-ray correlators, being carried out independently but concurrently.

\section{Four light-ray correlator for gluons}\label{sec:four-light-gluon}
Light-ray operators are naturally defined in a Lorentzian CFT which is achieved by working in a (2,2) signature bulk spacetime~\cite{Atanasov:2021oyu}. We start with the four-point gluon celestial amplitude in this spacetime, which was first computed in \cite{Hu:2022syq} (the analogous result in a (3,1) signature bulk spacetime was first computed in \cite{Pasterski:2017ylz})\footnote{Note that we are using a slightly different pre-factor in this paper. We can compare the two using the equation $\prod_{i<j} \left|z_{ij}\right|^{\frac{h}{3}-h_i-h_j}\left|\zb_{ij}\right|^{\frac{\bh}{3}-\bh_i-\bh_j} = X\left(z_{ij},\zb_{ij}\right) \left|z\right|^{\frac{h}{3}}\left|\zb\right|^{\frac{\bh}{3}} \left|1-z\right|^{\frac{h}{3}-h_2-h_3}\left|1-\zb\right|^{\frac{\bh}{3}-\bh_2-\bh_3}$.}
\begin{equation}
    \begin{split}
        \langle \mathcal{O}{}_{\Delta_1,-}(z_1,\bar{z}_1)
			&\mathcal{O}{}_{\Delta_2,-}(z_2,\bar{z}_2)\,\mathcal{O}_{\Delta_3,+}(z_3,\bar{z}_3)\mathcal{O}_{\Delta_4,+}(z_4,\bar{z}_4)\rangle\\
		&=~ \pi\,\d(\beta)\delta(z-\bar{z}){\rm sgn}\left(\frac{z}{z-1}\right)\left|z\right|^3 \left|1-z\right|^{1-\D_2-\D_3} X\left(z_{ij},\zb_{ij}\right)~,
    \end{split}
    \label{eq:4ptcorrelatorgluons}
\end{equation}
where
\begin{equation}
\resizebox{0.9\textwidth}{!}{$%
    \begin{aligned}
         X(z_{ij},\bz_{ij}) ~=&~  \frac{1}{|z_{12}|^{h_1+h_2}|z_{34}|^{h_3+h_4}}\left|\frac{z_{24}}{z_{14}}\right|^{h_{12}}\left|\frac{z_{14}}{z_{13}}\right|^{h_{34}}\frac{1}{|\bar{z}_{12}|^{\bar{h}_1+\bar{h}_2}|\bar{z}_{34}|^{\bar{h}_3+\bar{h}_4}}\left|\frac{\bar{z}_{24}}{\bar{z}_{14}}\right|^{\bar{h}_{12}}\left|\frac{\bar{z}_{14}}{\bar{z}_{13}}\right|^{\bar{h}_{34}}~,\\
    h_i &~=~ \frac{\D_i+J_i}{2} ~~,~~ \bar{h}_i ~=~ \frac{\D_i-J_i}{2}~~,~~
    h_{ij} ~=~ h_i - h_j ~~,~~ \bar{h}_{ij} ~=~ \bar{h}_i - \bar{h}_j ~~,\\
   &~ z_{ij} ~=~ z_i - z_j~~,~~ \zb_{ij} = \zb_i - \zb_j~~,~~ z = \frac{z_{12}z_{34}}{z_{13}z_{24}} ~~,~~ \\
   &~J_1=J_2=-J_3=-J_4=-1~~, \quad \textrm{and} \quad \beta = \sum_{i=1}^4\Delta_i - 4~~.
    \end{aligned}$}%
\end{equation}

Light-ray operators are defined by\footnote{As in \cite{Hu:2022syq}, we note that this definition is appropriate for the situation in this paper where ${\cal O} \sim {\cal O}^{\text{in}} + {\cal O}^{\text{out}}$.}
\begin{equation}
    \begin{split}
        &{\bf L} [\mathcal{O}_{\Delta,J}](z,\bar{z}) := \, \int_{-\infty}^{\infty} \, \frac{d z^{\prime}}{|z^{\prime}-z|^{2-\Delta-J}} \, \mathcal{O}_{\Delta,J}(z^{\prime},\bar{z})~, \\
&\bar{\bf L}[\mathcal{O}_{\Delta,J}](z,\bar{z}) := \, \int_{-\infty}^{\infty} \, \frac{d \bar{z}^{\prime}}{|\bar{z}^{\prime}-\bar{z}|^{2-\Delta+J}} \, \mathcal{O}_{\Delta,J}(z,\bar{z}^{\prime})~.
    \end{split}
\end{equation}
We will refer to these as the ``holomorphic" and ``anti-holomorphic" light-ray operators respectively. Note that the conformal weights of ${\bf L} [\mathcal{O}_{\Delta,J}]$ are $\left(1-\frac{\D+J}{2}, \frac{\D-J}{2}\right)$ and those of $\bar{\bf L}[\mathcal{O}_{\Delta,J}]$ are $\left(\frac{\D+J}{2},1-\frac{\D-J}{2}\right)$. %Consequently, a correlator involving them transforms with these weights. We can compute correlation functions involving light-ray operators by integrating (\ref{eq:arbitrary4ptcorrelator}) along appropriate contours. 

In this section, we will compute the correlator involving two holomorphic and two anti-holomorphic light-ray operators 
\begin{align}
&\langle \bar{\bf L}[\mathcal{O}_{\Delta_1,-}](z_1,\bar{z}_1) \,\bar{\bf L}[\mathcal{O}_{\Delta_2,-}](z_2,\bar{z}_2)\,{\bf L}[\mathcal{O}_{\Delta_3,+}](z_3,\bar{z}_3)\,{\bf L}[\mathcal{O}_{\Delta_4,+}](z_4,\bar{z}_4) \rangle \nonumber \\
&=\int_{-\infty}^{\infty} \, \frac{d \zb_1^{\prime}}{|\zb_1^{\prime}-\zb_1|^{1-\Delta_1}}\frac{d \zb_2^{\prime}}{|\zb_2^{\prime}-\zb_2|^{1-\Delta_2}}\frac{d z_3^{\prime}}{|z_3^{\prime}-z_3|^{1-\Delta_3}}\frac{d z_4^{\prime}}{|z_4^{\prime}-z_4|^{1-\Delta_4}} \, \\
&\qquad\qquad\times  \langle \mathcal{O}{}_{\Delta_1,-}(z_1,\bar{z}'_1)
			\mathcal{O}{}_{\Delta_2,-}(z_2,\bar{z}'_2)\,\mathcal{O}_{\Delta_3,+}(z'_3,\bar{z}_3)\mathcal{O}_{\Delta_4,+}(z'_4,\bar{z}_4)\rangle~.\nonumber
\end{align}
After performing the $\zb_1^{\prime}$ integral using the delta function in (\ref{eq:4ptcorrelatorgluons}) and changing variables to 
\begin{align}
t~=~\frac{\bar{z}_{12^{\prime}}\,\bar{z}_{34}}{\bar{z}_{13}\,\bar{z}_{2^\prime4}}~,\quad x~=~\frac{z_{12}\,z_{3^\prime4^\prime}}{z_{13^\prime}\,z_{24^\prime}}~, \quad y~=~\frac{z_{12}\,z_{34^\prime}}{z_{13}\,z_{24^\prime}} ~,
\end{align}
we end up with 
\begin{equation}
    \begin{split}
        &\langle \bar{\bf L}[\mathcal{O}_{\Delta_1,-}](z_1,\bar{z}_1) \,\bar{\bf L}[\mathcal{O}_{\Delta_2,-}](z_2,\bar{z}_2)\,{\bf L}[\mathcal{O}_{\Delta_3,+}](z_3,\bar{z}_3)\,{\bf L}[\mathcal{O}_{\Delta_4,+}](z_4,\bar{z}_4) \rangle \\
~&=~ \pi\, \delta(\beta)\, X(z_{ij},\bz_{ij})\,\left|\frac{z}{\bz}\right|^{\frac{\D_1+\D_2-2}{2}}\,|1-z|^{2-\D_2-\D_4} \,\mathcal{G}(z,\bar{z})~~.
    \end{split}
    \label{equ:four-light-setup}
\end{equation}
The pre-factor $X$ now involves the conformal weights of the light-ray operators which are
\begin{equation}
\begin{split}
   & (h_1,\bar{h}_1)~=~\left(\frac{\D_1-1}{2},\frac{1-\D_1}{2}\right)~~,~~(h_2,\bar{h}_2)~=~\left(\frac{\D_2-1}{2},\frac{1-\D_2}{2}\right)~~,~~\\
    &(h_3,\bar{h}_3)~=~\left(\frac{1-\D_3}{2},\frac{\D_3-1}{2}\right)~~,~~(h_4,\bar{h}_4)~=~\left(\frac{1-\D_4}{2},\frac{\D_4-1}{2}\right)~~,~~
    \end{split}
\end{equation} 
and 
\begin{equation}
    \begin{split}
        \mathcal{G}(z,\bar{z}) ~:=~ \int_{-\infty}^{\infty} dt\,dx\,dy\,  
\frac{1}{x\left(x-1\right)} &|\bar{z}-t|^{\Delta_2-1}\,|y-z|^{\Delta_4-1}\,|1-t|^{\Delta_4-2}\,|x-t|^{\Delta_1-1}  \\
& \times
|x-y|^{\Delta_3-1}\,|y-1|^{\Delta_2-2}~.
    \end{split}
    \label{4ptlightcorr}
\end{equation}
Our strategy to compute this integral is to first reduce it to an integral over the product of the following four-marked point integrals:
\begin{align}
\mathcal{F}_1(z,x) &= \int_{-\infty}^{\infty} \, dy \, |y-z|^{\Delta_4-1} |x-y|^{\Delta_3-1} |y-1|^{\Delta_2-2} ~,\\
\mathcal{F}_2(x,\bar{z}) &=\int_{-\infty}^{\infty} \, dt \, |x-t|^{\Delta_1-1} |1-t|^{\Delta_4-2} |\bar{z}-t|^{\Delta_2-1}~.
\end{align}
These integrals were first encountered in the evaluation of correlators with two light-ray operators in \cite{Hu:2022syq} and can be expressed in terms of Gauss hypergeometric functions. For a complete description of these results, we refer the reader to Appendix \ref{appen:integral-gluon}. The desired form for $\mathcal{G}(z,\bar{z})$ is
\begin{align}
\mathcal{G}(z,\bar{z}) = \int_{-\infty}^{\infty} \, \frac{dx}{x(x-1)} \,
 \mathcal{F}_1(z,x) \, \mathcal{F}_2(x,\bar{z})~.
\label{condensedfourlighttrans}
\end{align}
For simplicity, we will focus on the case $z,\bz \in [0,1]$ for which the $x$ integral splits into two regions:
\begin{equation}
    \begin{split}
        {\cal G}(z,\bz)
        ~&=~ {\cal G}(z,\bz)\bigg|_{\textrm{I}} + {\cal G}(z,\bz)\bigg|_{\textrm{II}} \\
        ~&=~ \int_{-\infty}^1\,dx\,\frac{1}{x(x-1)}\,
        {\cal F}_1(z,x)|_{x,z<1}\,{\cal F}_2(x,\bz)|_{x,\bz<1}\\
      ~&\qquad+~ \int_1^{\infty}\,dx\,\frac{1}{x(x-1)}\,
       {\cal F}_1(z,x)|_{z<1<x}\,{\cal F}_2(x,\bz)|_{\bz<1<x}~.
    \end{split}
    \label{equ:G-two-regions}
\end{equation}
The integrals can be explicitly evaluated in terms of Fox H-functions \cite{Hfunc1,Hfunc2,Hfunc3,Hfunc4,multiHfunc} and I-functions \cite{Ifunc1, Ifunc2} which are special functions belonging to a general class of Mellin-Barnes integrals. Here, we flesh out the evaluation of $\mathcal{G}(z,\bz)\big|_{\textrm{II}}$ and refer the reader to Appendix \ref{appen:I-func} for more details of these special functions and to Appendix \ref{appen:g1} for the evaluation of $\mathcal{G}(z,\bz)\big|_{\textrm{I}}$ (the details heavily rely on the ordering we choose for $x,z,$ and $\bz$). 

\subsection{Evaluation of \texorpdfstring{${\cal G}(z,\bz)\big|_{\textrm{II}}$}{TEXT}} \label{evalutionoffourlighttransforms}
Evaluating \gii using the definitions of ${\cal F}_1(z,x)$ and ${\cal F}_2(x,\bz)$ in (\ref{equ:tildeF-II}) and (\ref{equ:F-II}) results in a sum of four integrals, one of which is  
\begin{equation}
    \begin{split}
        \text{\gii} \supset \int_{1}^{\infty} \frac{dx}{x\left(x-1\right)}& \left(1-\zb\right)^{1-\D_3}\left(1-z\right)^{1-\D_1} C\left(\D_4,\D_1-1\right)\, C\left(\D_2,\D_3-1\right)\\
    & \times \hypf\left(1-\D_3,\D_1-1,\D_4+\D_1-1,\frac{1-x}{1-z}\right)\\ 
    & \qquad\times\hypf\left(1-\D_1,\D_3-1,\D_2+\D_3-1,\frac{1-x}{1-\zb}\right),
    \end{split}
\end{equation}
where we've eliminated the absolute value signs keeping in mind that we're working in the region $0<z,\zb<1<x$. We then use the Mellin-Barnes representation of the Gauss hypergeometric function
\begin{align}
_2F_1[a,b,c;-|z|] = \frac{1}{2\pi i}\frac{\Gamma(c)}{\Gamma(a) \Gamma(b)} \, \int_{-i \infty}^{i\infty} \, ds \, \frac{\Gamma(-s) \Gamma(a+s) \Gamma(b+s)}{\Gamma(c+s)} \, |z|^{s}~,
\label{mellinbarnesgauss}
\end{align}
to obtain 
\begin{equation}
\resizebox{0.9\textwidth}{!}{$%
    \begin{aligned}
        \text{\gii} &\supset \frac{1}{\left(2\pi i\right)^2} \left(1-\zb\right)^{1-\D_3}\left(1-z\right)^{1-\D_1} C\left(\D_4,\D_1-1\right)\, C\left(\D_2,\D_3-1\right)\\
    & \times\int_{-i \infty}^{i\infty} \, ds_1  \frac{\Gamma\left(\D_4+\D_1-1\right)}{\Gamma\left(1-\D_3\right)\Gamma\left(\D_1-1\right)}\frac{\Gamma\left(-s_1\right)\Gamma\left(1-\D_3+s_1\right)\Gamma\left(\D_1-1+s_1\right)}{\Gamma\left(\D_4+\D_1-1+s_1\right)} \left(1-z\right)^{-s_1}\\
    & \times \int_{-i \infty}^{i\infty} \, ds_2 \frac{\Gamma\left(\D_2+\D_3-1\right)}{\Gamma\left(1-\D_1\right)\Gamma\left(\D_3-1\right)}\frac{\Gamma\left(-s_2\right)\Gamma\left(1-\D_1+s_2\right)\Gamma\left(\D_3-1+s_2\right)}{\Gamma\left(\D_2+\D_3-1+s_2\right)}\left(1-\zb\right)^{-s_2}\\
    & \times\int_1^\infty dx\,x^{-1} \left(x-1\right)^{-1+s_1+s_2}~.
    \end{aligned}$}%
\end{equation}
The $x$-integral evaluates to $B\left(s_1+s_2,1-s_1-s_2\right)$ and comparing with equations (\ref{foxhfunc}), (\ref{gammasofhfunc}), we can identify the residual double Mellin-Barnes to be
\begin{equation}
\resizebox{0.9\textwidth}{!}{$%
    \begin{aligned}
        &\frac{1}{\left(2\pi i\right)^2}\int_{-i \infty}^{i\infty} \, ds_1 \,ds_2\, \left[\Gamma\left(1-s_1-s_2\right)\Gamma\left(s_1+s_2\right)\right] \times\left[\frac{\Gamma\left(-s_1\right)\Gamma\left(1-\D_3+s_1\right)\Gamma\left(\D_1-1+s_1\right)}{\Gamma\left(\D_4+\D_1-1+s_1\right)}\right]\\
    & \qquad\qquad \times \left[\frac{\Gamma\left(-s_2\right)\Gamma\left(1-\D_1+s_2\right)\Gamma\left(\D_3-1+s_2\right)}{\Gamma\left(\D_2+\D_3-1+s_2\right)}\right] \times \left(1-\zb\right)^{-s_2}\left(1-z\right)^{-s_1}\\
    &=\textrm{H}^{1,1:\,2,1;\,2,1}_{1,1:\,2,2;\,2,2} \left(
\begin{array}{c}
1-z \\ 1-\bar{z}
\end{array}\middle\vert
\begin{array}{c}
(0,1):\,(1,\Delta_1+\Delta_4-1,1,1);\, (1,\Delta_2+\Delta_3-1,1,1)\\
(0,1):\,(1-\Delta_3,\Delta_1-1,1,1);\,(1-\Delta_1,\Delta_3-1,1,1)\\
\end{array}
\right)~.
    \end{aligned}$}%
\end{equation}
The remaining three integrals can be performed in a similar manner. The final result is thus a sum of four Fox H-functions as seen below:
\begingroup
\allowdisplaybreaks
\begin{align}
%\resizebox{0.9\textwidth}{!}{$%
%    \begin{aligned}
        &\mathcal{G}(z,\bz)\big|_{\textrm{II}} ~=~ \int_{1}^{\infty} \frac{dx}{x(x-1)} \, \mathcal{F}_1(z,x)\bigg|_{z<1<x} \, \mathcal{F}_2(x,\bar{z})\bigg|_{\bar{z}<1<x} \nonumber\\
&= |1-z|^{1-\Delta_1} |1-\bar{z}|^{1-\Delta_3} C(\Delta_4,\Delta_1-1) C(\Delta_2,\Delta_3-1) \frac{\Gamma(\Delta_2+\Delta_3-1) \Gamma(\Delta_1+\Delta_4-1)}{\Gamma(1-\Delta_1) \Gamma(\Delta_3-1) \Gamma(1-\Delta_3) \Gamma(\Delta_1-1)}  \nonumber\\
&~~~~\times \textrm{H}^{1,1:\,2,1;\,2,1}_{1,1:\,2,2;\,2,2} \left(
\begin{array}{c}
1-z \\ 1-\bar{z}
\end{array}\middle\vert
\begin{array}{c}
(0,1):\,(1,\Delta_1+\Delta_4-1,1,1);\, (1,\Delta_2+\Delta_3-1,1,1) \\
(0,1):\,(1-\Delta_3,\Delta_1-1,1,1);\,(1-\Delta_1,\Delta_3-1,1,1)\\
\end{array}
\right)~  \nonumber\\
&+ \bigg\{ |1-z|^{1-\Delta_1} |1-\bar{z}|^{\Delta_2-1} C(\Delta_4,\Delta_1-1) C(\Delta_1,\Delta_4-1) \frac{ (\Gamma(\Delta_1+\Delta_4-1))^2}{\Gamma(1-\Delta_3) \Gamma(\Delta_1-1) \Gamma(1-\Delta_2) \Gamma(\Delta_4-1)} \nonumber \\
&~~~~\times \textrm{H}^{1,1:\,2,1;\,2,1}_{1,1:\,2,2;\,2,2} \left(
\begin{array}{c}
1-z \\ 1-\bar{z}
\end{array}\middle\vert
\begin{array}{c}
(\Delta_1+\Delta_4-2,1):\,(1,\Delta_1+\Delta_4-1,1,1);\, (1,\Delta_1+\Delta_4-1,1,1)\\
(\Delta_1+\Delta_4-2,1):\,(1-\Delta_3,\Delta_1-1,1,1);\,(1-\Delta_2,\Delta_4-1,1,1)\\
\end{array}
\right)  \nonumber\\
&~~~~+ \big(z \leftrightarrow \bar{z}, \Delta_1 \leftrightarrow \Delta_3, \Delta_2 \leftrightarrow \Delta_4\big)~\bigg\}  \nonumber\\
&+ |1-z|^{\Delta_4-1} |1-\bar{z}|^{\Delta_2-1} \, C(\Delta_3,\Delta_2-1) \, C(\Delta_1,\Delta_4-1) \frac{\Gamma(\Delta_2+\Delta_3-1) \Gamma(\Delta_1+\Delta_4-1)}{\Gamma(1-\Delta_4) \Gamma(\Delta_2-1) \Gamma(1-\Delta_2) \Gamma(\Delta_4-1)}  \nonumber\\
&~~~~\times \textrm{H}^{1,1:\,2,1;\,2,1}_{1,1:\,2,2;\,2,2} \left(
\begin{array}{c}
1-z \\ 1-\bar{z}
\end{array}\middle\vert
\begin{array}{c}
(0,1):\,(1,\Delta_2+\Delta_3-1,1,1);\, (1,\Delta_1+\Delta_4-1,1,1)\\
(0,1):\,(1-\Delta_4,\Delta_2-1,1,1);\,(1-\Delta_2,\Delta_4-1,1,1)\\
\end{array}
\right)~.
%    \end{aligned}
%$}%
    \label{secondregionfinal}
\end{align}
\endgroup
The evaluation of \gi is more involved due to the presence of additional singular points at $x=z,\zb$ in the $x$-integral. The $x$-integrals generically evaluate to an Appell function $F_1$ and contributes an additional Mellin-Barnes integral. The resulting special function is a generalization of the Fox H-function, called the I-function. We relegate the details of this computation to Appendix \ref{appen:g1}.

\subsection{Conformal block decomposition}
Note that as shown in (\ref{condensedfourlighttrans}), the cross ratios $z$ and $\bz$ only occur in the hypergeometric functions. For each ${\cal F}_1$ and ${\cal F}_2$, we know how to do the conformal block decompositions for the simplest gluon exchange term, namely the first terms in both (\ref{equ:tildeF-I}) and (\ref{equ:F-I}) based on the results in \cite{Hu:2022syq}. Therefore for the corresponding term, we can read off the conformal block decomposition result out of the four terms in the first integral in (\ref{equ:G-two-regions}). The spectrum of exchanged states is
\begin{equation}
    (h,\bar{h}) ~=~ \left(\,\frac{\D_1+\D_2}{2}-1+m,\frac{\D_3+\D_4}{2}-1+n \,\right)~~~m,n\in \mathbb{Z}_{\ge 0}~.
\end{equation}

\section{Light-ray correlators for scalars}\label{sec:scalar-light-transf}

In this section, we consider the scattering of four massless scalars mediated by a massive one. Throughout this section, we will drop the spin label on the CCFT operators. The amplitude is \footnote{This amplitude was also considered in \cite{Atanasov:2021cje, Nandan:2019jas} where the authors restricted themselves to different channels in order to analyze the conformal block decomposition. Restrictions of this sort are not feasible for the purpose of computing light transforms in the formalism of our paper.}
\begin{align}
     \label{equ:scalar-amp-mom}
     \mathbf{M}\left(s,t\right)
      = g^2\,\left( \frac{1}{m^2 -s} ~+~ \frac{1}{m^2 -t} ~+~ \frac{1}{m^2+s+t} \right)~.
\end{align}
where 
\begin{equation}
        s = -4\,\varepsilon_3\varepsilon_4\,\o_3\o_4\,z_{34}\bar{z}_{34} \quad , \quad t = -4\,\varepsilon_2\varepsilon_4\,\o_2\o_4\,z_{24}\bar{z}_{24}~, 
\end{equation}
The analog of (\ref{eq:4ptcorrelatorgluons}) for this theory is
\begin{equation}\resizebox{0.9\textwidth}{!}{$%
    \begin{aligned}
        &\langle\,{\cal O}_{\D_1}(z_1,\bar{z}_1){\cal O}_{\D_2}(z_2,\bar{z}_2){\cal O}_{\D_3}(z_3,\bar{z}_3){\cal O}_{\D_4}(z_4,\bar{z}_4)\,\rangle\\
         ~=&~ N(\b)\,|z(1-z)|^{\frac{\b}{2}+2}\,|1-z|^{-\D_2-\D_3}\,\d(z-\bar{z})\,X(z_{ij},\bz_{ij})\,\Bigg\{\,[{\rm sgn}((z-1)z_{34}z_{23}z_{24}){\rm sgn}(\bz_{23}\bz_{24}\bz_{34})]^{-\hb}|z|^{-\hb}\\
        & ~+~ [{\rm sgn}((1-z)zz_{12}z_{13}z_{23}){\rm sgn}(\bz_{12}\bz_{13}\bz_{23}) ]^{-\hb} ~+~ [{\rm sgn}(-z z_{12}z_{14}z_{24}){\rm sgn}(\bz_{12}\bz_{14}\bz_{24})]^{-\hb}|1-z|^{-\hb} \,\Bigg\}~,
    \end{aligned}$}%
    \label{equ:four-scalar-final}
\end{equation}
where 
\begin{align}
    N(\b) ~=&~ 2^{-\b-2}\,g^2\,\pi\,|m|^{\b-2}\,\csc{\frac{\b\pi}{2}}~~.
\end{align}
The presence of the mass term leads to residual sgn functions even after the summation over $\epsilon_i$. This impacts the computation of light-ray correlators. In the limit $m\to\infty$ with $g/m$ held fixed, the exchange interaction effectively reduces to a contact interaction and $N(\b) \to \d(\b)$ as pointed out in \cite{Atanasov:2021cje}. In particular, the sgn functions vanish in this limit and the correlator resembles that of gluons.

\subsection{Two light-ray correlator}\label{sec:double-light-scalar}
Owing to the differences cited above, the correlator involving two light-ray operators differs from that of gluons. This correlator is also essential in computing the four light-ray correlator. The final result can still be expressed in terms of Gauss hypergeometric functions albeit with more complicated coefficients. Starting with (\ref{equ:four-scalar-final}), we have
\begin{equation}
\resizebox{0.9\textwidth}{!}{$%
    \begin{aligned}
        &\langle\,{\bf \bar{L}}[{\cal O}_{\D_1}](z_1,\bar{z}_1){\bf \bar{L}}[{\cal O}_{\D_2}](z_2,\bar{z}_2){\cal O}_{\D_3}(z_3,\bar{z}_3){\cal O}_{\D_4}(z_4,\bar{z}_4)\,\rangle\\
        ~=&~ \int_{-\infty}^{\infty}\,\frac{d\bar{z}_1'}{|\bar{z}_1'-\bar{z}_1|^{2-\Delta_1}}\frac{d\bar{z}_2'}{|\bar{z}_2'-\bar{z}_2|^{2-\Delta_2}}\,\langle\,{\cal O}_{\D_1}(z_1,\bar{z}'_1){\cal O}_{\D_2}(z_2,\bar{z}'_2){\cal O}_{\D_3}(z_3,\bar{z}_3){\cal O}_{\D_4}(z_4,\bar{z}_4)\,\rangle \\
        ~=&~ N(\b)\,X(z_{ij},\bz_{ij})\,|\bz|^{\frac{\D_3+\D_4}{2}-\hb}\,|z|^{\frac{\D_3+\D_4}{2}}\, {\cal F}_{\phi}(z,\bar{z})~,
    \end{aligned}$}%
    \label{equ:scalar-LLOO}
\end{equation}
where  ${\cal F}_{\phi}(z,\bar{z})$ is again an integral over four marked points:
\begin{equation}
    \begin{split}
        {\cal F}_{\phi}(z,\bar{z}) ~&=~ \int_{-\infty}^{\infty}\,dt\,|z-t|^{\D_1-2}\,|\bar{z}-t|^{\D_2-2}\,|1-t|^{\D_4-1-\frac{\b}{2}}\\
        &\quad \times~ \Bigg\{\,[{\rm sgn}((z-1)z_{34}z_{23}z_{24}){\rm sgn}(\bz_{13}\bz_{14}\bz_{34}(1-t))]^{-\hb}|z|^{-\hb}\\
        &\qquad ~~+~ [{\rm sgn}((1-z)z_{12}z_{13}z_{23}){\rm sgn}(\bz_{13}\bz_{14}\bz_{34}(1-t)) ]^{-\hb}\\
        &\qquad 
        ~~+~ [{\rm sgn}((z-1) z_{12}z_{14}z_{24}){\rm sgn}(\bz_{13}\bz_{14}\bz_{34}(1-t))]^{-\hb}|1-z|^{-\hb} \,\Bigg\}\\
        &=~ {\cal F}_{\phi,1}(z,\bar{z})+{\cal F}_{\phi,2}(z,\bar{z})+{\cal F}_{\phi,3}(z,\bar{z})~.
    \end{split}
    \label{eq:4markedptscalar}
\end{equation}
$X(z_{ij},\bz_{ij})$ is the suitable pre-factor for the correlator which has conformal weights
\begin{equation}
\begin{split}
   & (h_1,\bar{h}_1)=\left(\frac{\D_1}{2},1-\frac{\D_1}{2}\right)~~,~~(h_2,\bar{h}_2)=\left(\frac{\D_2}{2},1-\frac{\D_2}{2}\right)~~,~~\\%h_{ij}=h_i-h_j~~,\\
    &(h_3,\bar{h}_3)=\left(\frac{\D_3}{2},\frac{\D_3}{2}\right)~~~~,~~~~(h_4,\bar{h}_4)=\left(\frac{\D_4}{2},\frac{\D_4}{2}\right)~.%~~~~,~~~~\bh_{ij}=\bh_i-\bh_j~~,.
    \end{split}
\end{equation}
The s-, t-, and u-channels (in the bulk) contribute different sign functions to the integral (\ref{eq:4markedptscalar}). These eventually lead to different phases in the coefficient of the  Gauss hypergeometric functions. The integral can be evaluated by using techniques similar to those of \cite{Hu:2022syq} and for $0<z<\zb<1$, we find
\begin{equation}
    \begin{split}
    {\cal F}_{\phi,i}(z,\bz)|_{z,\bz<1\,\cup\,z,\bz>1} 
         ~=&~  {\cal S}_i(z_{ab})\,|1-z|^{\hb-\D_3}\, C^{(1)}\left(\D_4-\hb,\D_3-\hb\right)\\
         &\qquad \times \,   _2F_1\left(2-\D_2,\D_3-\hb,\D_3+\D_4-\b,\frac{\bz-z}{1-z}\right)\\
        +&~ {\cal S}_i(z_{ab})\,|\bz-z|^{\D_1+\D_2-3}\,|1-z|^{\D_4-1-\frac{\b}{2}}\,C\left(\D_1-1,\D_2-1\right)\\
        &\qquad \times \,  _2F_1\left(\D_1-1,1+\hb-\D_4,\D_1+\D_2-2,\frac{\bz-z}{1-z}\right)~,\\
    \end{split}
    \label{equ:F-phi-I}
\end{equation}
where
\begin{equation}
    C^{(1)}(a,b) ~:=~ e^{-\hb i \pi}  B(a,b) + B(a,1-a-b) + B(b,1-a-b)
\end{equation}
and 
\begin{align}
{\cal S}_i(z_{ab}) ~=~    \begin{cases}
        [-{\rm sgn}(z_{34}z_{23}z_{24}){\rm sgn}(\bz_{13}\bz_{14}\bz_{34})]^{-\hb}|z|^{-\hb} \qquad &i = 1\\
       [{\rm sgn}(z_{12}z_{13}z_{23}){\rm sgn}(\bz_{13}\bz_{14}\bz_{34}) ]^{-\hb}\qquad &i = 2\\
       [-{\rm sgn}( z_{12}z_{14}z_{24}){\rm sgn}(\bz_{13}\bz_{14}\bz_{34})]^{-\hb}|1-z|^{-\hb}\qquad &i = 3
    \end{cases}
\end{align}
The phase factor $(-1)^{-\hb} = e^{-\hb i \pi}$ comes from the sign functions. The result for the other regions can be found in Appendix \ref{appen:integral-scalar}.

\subsection{Four light-ray correlator}
The computation of the correlator involving four scalar light-ray operators begins in a manner analogous to Section [\ref{sec:four-light-gluon}]. 
\begin{equation}
    \begin{split}
    &\langle\,{\bf \bar{L}}[{\cal O}_{\D_1}](z_1,\bar{z}_1){\bf \bar{L}}[{\cal O}_{\D_2}](z_2,\bar{z}_2){\bf {L}}[{\cal O}_{\D_3}](z_3,\bar{z}_3){\bf {L}}[{\cal O}_{\D_4}](z_4,\bar{z}_4)\,\rangle \\
        %%%%%%%%%%%%%%%%%%%%%%%%%%%%%%%%%%%%%%%%%%%%%%%%%
        %%%%%%%%%%%%%%%%%%%%%%%%%%%%%%%%%%%%%%%%%%%%%%%%%
         ~&=~ N(\b)\,|\bz|^{\frac{\D_3+\D_4}{2}-\hb}\,|z|^{2-\frac{\D_3+\D_4}{2}}|1-z|^{\frac{\D_1+\D_3-\D_2-\D_4}{2}}\,X(z_{ij},\bz_{ij}) \\
         &~~\times~ \int_{-\infty}^{\infty}\,dx\,\int_{-\infty}^{\infty}\,dy\,|1-y|^{\D_2-1-\hb}\,|z-y|^{\D_4-2}|x-y|^{\D_3-2}\\
         &~~\times~ \int_{-\infty}^{\infty}\,dt\,
         %%%%%%%%%%%%%%%%%%%%%%%%%%%%%%%%%%%%%%%%%%%%%%%%%
         |x-t|^{\D_1-2}\,|\bar{z}-t|^{\D_2-2}\,|1-t|^{\D_4-1-\frac{\b}{2}}\, S_{\phi}(x,y,t)\\
    \end{split}
    \label{equ:scalar-LbLbLL}
\end{equation}
where 
\begin{equation}\resizebox{0.9\textwidth}{!}{$%
    \begin{aligned}
      S_{\phi}(x,y,t)  ~=&~  \Bigg\{\,[{\rm sgn}(x(y-1)(1-t)){\rm sgn}(z_{12}z_{13}z_{23}){\rm sgn}(\bz_{13}\bz_{14}\bz_{34})]^{-\hb}|x|^{-\hb}\\
        &\qquad ~~+~ [{\rm sgn}((1-y)(1-t)){\rm sgn}(z_{12}z_{13}z_{23}){\rm sgn}(\bz_{13}\bz_{14}\bz_{34}) ]^{-\hb}\\
        &\qquad 
        ~~+~ [{\rm sgn}((x-1)(1-y)(1-t)){\rm sgn}( z_{12}z_{13}z_{23}){\rm sgn}(\bz_{13}\bz_{14}\bz_{34})]^{-\hb}|1-x|^{-\hb} \,\Bigg\}~,\\
    \end{aligned}$}%
\end{equation}
and the conformal weights in the $X(z_{ij},\bz_{ij})$ function are
\begin{equation}
\begin{split}
   & (h_1,\bar{h}_1)=\left(\frac{\D_1}{2},1-\frac{\D_1}{2}\right)~~,~~(h_2,\bar{h}_2)=\left(\frac{\D_2}{2},1-\frac{\D_2}{2}\right)~~,~~\\%h_{ij}=h_i-h_j~~,\\
    &(h_3,\bar{h}_3)=\left(1-\frac{\D_3}{2},\frac{\D_3}{2}\right)~~,~~(h_4,\bar{h}_4)=\left(1-\frac{\D_4}{2},\frac{\D_4}{2}\right)~~.~~%\bh_{ij}=\bh_i-\bh_j~~.
    \end{split}
\end{equation} 
The sgn functions hinder the factorization of this integral and we cannot express this in a form similar to (\ref{condensedfourlighttrans}). However, we can proceed by breaking up the integral into regions where they have fixed signs and then perform the integral using tricks similar to Section [\ref{sec:four-light-gluon}]. The final result is then expressible in terms of Fox H-functions and I-functions, as in the gluon case.

%-------------------------------------------------------------------------	

\acknowledgments
	We are grateful to Rishabh Bhardwaj, Luke Lippstreu, Lecheng Ren and Marcus Spradlin for useful
	comments and discussion.
	This work was supported in part by the US Department of Energy under contract {DE}-{SC}0010010 Task A and by Simons Investigator Award \#376208.
	The research of Y. Hu is supported in part by the endowment from the Ford Foundation Professorship of Physics and the Physics Dissertation Fellowship provided by the Department of Physics at Brown University. Y. Hu also acknowledges the support of the Brown Theoretical Physics Center.

%----------------------------------------------------------------------------------	
	\appendix
%---------------------------------------------------------------------------------------	

\section{The library of four-marked-point integrals}\label{appen:four-marked-point-integrals}
This appendix is devoted to tabulating all four-marked-point integrals encountered in the gluon and scalar light transform calculations. 

\subsection{Gluon integrals}\label{appen:integral-gluon}

\begin{equation}
    \begin{split}
       & {\cal F}_1(z,x)|_{z,x<1 \, \cup\, z,x>1} \\
        &~~= |x-1|^{1-\Delta_1}\, _2F_1\left(1-\Delta_4,\Delta_1-1,\Delta_1+\Delta_2-2,\frac{x-z}{x-1}\right)\,C(\Delta_2-1,\Delta_1-1)\\
            &~~+|x-1|^{\Delta_2-2}|z-x|^{\Delta_3+\Delta_4-1}\,_2F_1\left(2-\Delta_2,\Delta_3,\Delta_3+\Delta_4,\frac{x-z}{x-1}\right)\,C(\Delta_3,\Delta_4)
    \end{split}
    \label{equ:tildeF-I}
\end{equation}

\begin{equation}
\resizebox{0.9\textwidth}{!}{$%
    \begin{aligned}
       & {\cal F}_1(z,x)|_{z<1<x\,\cup\,z>1>x} \\
      &~~= |1-z|^{1-\Delta_1}\,_2F_1\left(1-\Delta_3,\Delta_1-1,\Delta_4+\Delta_1-1,-\frac{|x-1|}{|1-{z}|}\right)\,C(\Delta_4,\Delta_1-1)\\
        &~~+ |x-1|^{\Delta_3+\Delta_2-2}|1-{z}|^{\Delta_4-1}\,_2F_1\left(1-\Delta_4,\Delta_2-1,\Delta_3+\Delta_2-1,-\frac{|x-1|}{|1-{z}|}\right)\,C(\Delta_3,\Delta_2-1)
    \end{aligned}$}%
     \label{equ:tildeF-II}
\end{equation}

\begin{equation}
    \begin{split}
       & {\cal F}_2(x,\bz)|_{\bz,x<1\,\cup\,\bz,x>1} \\
        &~~=  |x-1|^{1-\Delta_3}\, _2F_1\left(1-\Delta_2,\Delta_3-1,\Delta_3+\Delta_4-2,\frac{\bar{z}-x}{1-x}\right)\,C(\Delta_4-1,\Delta_3-1)\\
            &~~+|x-1|^{\Delta_4-2}|\bar{z}-x|^{\Delta_1+\Delta_2-1}\,_2F_1\left(2-\Delta_4,\Delta_1,\Delta_1+\Delta_2,\frac{\bar{z}-x}{1-x}\right)\,C(\Delta_1,\Delta_2)
    \end{split}
     \label{equ:F-I}
\end{equation}

\begin{equation}\resizebox{0.9\textwidth}{!}{$%
    \begin{aligned}
       & {\cal F}_2(x,\bz)|_{\bz<1<x \, \cup\, \bz>1>x } \\
       &~~= |1-\bar{z}|^{1-\Delta_3}\,_2F_1\left(1-\Delta_1,\Delta_3-1,\Delta_2+\Delta_3-1,-\frac{|x-1|}{|1-\bar{z}|}\right)\,C(\Delta_2,\Delta_3-1)\\
        &~~+ |x-1|^{\Delta_1+\Delta_4-2}|1-\bar{z}|^{\Delta_2-1}\,_2F_1\left(1-\Delta_2,\Delta_4-1,\Delta_1+\Delta_4-1,-\frac{|x-1|}{|1-\bar{z}|}\right)\,C(\Delta_1,\Delta_4-1)
    \end{aligned}$}%
       \label{equ:F-II}
\end{equation}
where 
\begin{align}
	C(a,b)=B(a,b)+B(a,1-a-b)+B(b,1-a-b) \quad , \quad B(a,b) = \frac{\Gamma(a) \Gamma(b)}{\Gamma(a+b)}\,.
	\label{eq:Cdef}
\end{align}

\subsection{Scalar integrals} \label{appen:integral-scalar}
\begin{equation}\resizebox{0.9\textwidth}{!}{$%
    \begin{aligned}
    {\cal F}_{\phi,i}(z,\bz)|_{z,\bz<1\,\cup\,z,\bz>1} 
         ~=&~  {\cal S}_i(z_{ab})\,\left[|1-z|^{\hb-\D_3}\, C^{(1)}\left(\D_4-\hb,\D_3-\hb\right)\right.\\
         &\left.\qquad \qquad \times ~ _2F_1\left(2-\D_2,\D_3-\hb,\D_3+\D_4-\b,\frac{\bz-z}{1-z}\right)\right.\\
        +&~ \left.\,|\bz-z|^{\D_1+\D_2-3}\,|1-z|^{\D_4-1-\frac{\b}{2}}\,C\left(\D_1-1,\D_2-1\right)\right.\\
        &\left.\qquad \qquad \times ~  _2F_1\left(\D_1-1,1+\hb-\D_4,\D_1+\D_2-2,\frac{\bz-z}{1-z}\right)\right]\\
    \end{aligned}$}%
  %  \label{equ:F-phi-I}
\end{equation}
\begin{equation}\small
    \begin{split}
         {\cal F}_{\phi,i}(z,\bar{z})|_{z<1<\bz\,\cup\,z>1>\bz}  
         ~=&~ {\cal S}_i(z_{ab})\,\left[|\bz-1|^{\hb-\D_3}\,C^{(1,2)}\left(\D_2-1,\D_3-\hb\right)\right.\\
         &\left.\qquad\qquad \times\, _2F_1\left(\D_3-\hb,2-\D_1,\D_2+\D_3-1-\hb,-\frac{|1-z|}{|\bz-1|}\right) \right.\\
         +&\left. ~|1-z|^{\D_1+\D_4-2-\hb}|\bz-1|^{\D_2-2}\,C^{(3)}\left(\D_1-1,\D_4-\hb\right)\right.\\
         &\left.\qquad \qquad \times\, _2F_1\left(\D_4-\hb,2-\D_2,\D_1+\D_4-1-\hb,-\frac{|1-z|}{|\bz-1|}\right)\right]
    \end{split}
    \label{equ:F-phi-II}
\end{equation}
where
\begin{equation}
    \begin{split}
        C^{(1)}(a,b) ~:=&~ e^{-\hb i \pi}  B(a,b) + B(a,1-a-b) + B(b,1-a-b)\\
    C^{(1,2)}(a,b) ~:=&~ e^{-\hb i \pi} \,B(a,b) + e^{-\hb i \pi} \,B(a,1-a-b) + B(b,1-a-b) \\
    C^{(3)}(a,b)  ~:=&~B(a,b) + B(a,1-a-b) + e^{-\hb i \pi} \, B(b,1-a-b)
    \end{split}
\end{equation}
and 
\begin{align}
{\cal S}_i(z_{ab}) ~=~    \begin{cases}
        [-{\rm sgn}(z_{34}z_{23}z_{24}){\rm sgn}(\bz_{13}\bz_{14}\bz_{34})]^{-\hb}|z|^{-\hb} \qquad &i = 1\\
       [{\rm sgn}(z_{12}z_{13}z_{23}){\rm sgn}(\bz_{13}\bz_{14}\bz_{34}) ]^{-\hb}\qquad &i = 2\\
       [-{\rm sgn}( z_{12}z_{14}z_{24}){\rm sgn}(\bz_{13}\bz_{14}\bz_{34})]^{-\hb}|1-z|^{-\hb}\qquad &i = 3
    \end{cases}
\end{align}

\section{Definitions of Fox H-function and Generalized I-function}\label{appen:I-func}
Central to our discussion of the four-light transforms are special functions that belong to a general class of Mellin-Barnes type integrals, namely the Fox H-function \cite{Hfunc1,Hfunc2,Hfunc3,Hfunc4,multiHfunc} and the recently introduced generalized I-function \cite{Ifunc1,Ifunc2}. In this appendix, we briefly introduce the definitions of the bivariate Fox H-function $\textrm{H}(x,y)$ and the generalized I-function of $r$-variables $\textrm{I}(z_1,\dots,z_r)$. 

\paragraph{Fox H-function of two variables} The Fox H-function of two variables $\textrm{H}(x,y)$ (with the third characteristic)  is defined as
\begin{equation}
    \begin{split}
        \textrm{H}(x,y) &= \textrm{H}^{m_1,n_1:\,m_2,n_2;\,m_3,n_3}_{p_1,q_1:\,p_2,q_2;\,p_3,q_3}\left(
\begin{array}{c}
x\\y
\end{array}\middle\vert
\begin{array}{c}
(\alpha_{p_1}^{(1)}, a_{p_1}^{(1)}):\,(\alpha_{p_2}^{(2)}, a_{p_2}^{(2)});\, (\alpha_{p_3}^{(3)}, a_{p_3}^{(3)})\\
(\beta_{q_1}^{(1)}, b_{q_1}^{(1)}):\,(\beta_{q_2}^{(2)}, b_{q_2}^{(2)});\,(\beta_{q_3}^{(3)}, b_{q_3}^{(3)})\\
\end{array}
\right)  \\
& = \frac{1}{(2 \pi i)^2} \int_{-i \infty}^{+i \infty} \, ds\,  \int_{-i \infty}^{+i \infty} \, dt\, \phi_1(s+t) \, \phi_2(s) \, \phi_3(t) \, x^{-s} \, y^{-t} \,~, 
    \end{split}
    \label{foxhfunc}
\end{equation}
where
\begin{equation}
    \begin{split}
        \phi_1(s+t) &= \frac{\prod_{j=1}^{m_1}\Gamma\bigg(\beta_j^{(1)}+b_j^{(1)}(s+t)\bigg)\,\prod_{j=1}^{n_1}\Gamma\bigg(1-\alpha_j^{(1)}-a_j^{(1)}(s+t)\bigg)}{\prod_{j=n_1+1}^{p_1}\Gamma\bigg(\alpha_j^{(1)}+a_j^{(1)}(s+t)\bigg)\,\prod_{j=m_1+1}^{q_1}\Gamma\bigg(1-\beta_j^{(1)}-b_j^{(1)}(s+t)\bigg)}~,  \\
\phi_2(s) &= \frac{\prod_{j=1}^{m_2}\Gamma\bigg(\beta_j^{(2)}+b_j^{(2)}s\bigg)\,\prod_{j=1}^{n_2}\Gamma\bigg(1-\alpha_j^{(2)}-a_j^{(2)}s\bigg)}{\prod_{j=n_2+1}^{p_2}\Gamma\bigg(\alpha_j^{(2)}+a_j^{(2)}s\bigg)\,\prod_{j=m_2+1}^{q_2}\Gamma\bigg(1-\beta_j^{(2)}-b_j^{(2)}s\bigg)}~, \\
\phi_3(t) &= \frac{\prod_{j=1}^{m_3}\Gamma\bigg(\beta_j^{(3)}+b_j^{(3)}t\bigg)\,\prod_{j=1}^{n_3}\Gamma\bigg(1-\alpha_j^{(3)}-a_j^{(3)}t\bigg)}{\prod_{j=n_3+1}^{p_3}\Gamma\bigg(\alpha_j^{(3)}+a_j^{(3)}t\bigg)\,\prod_{j=m_3+1}^{q_3}\Gamma\bigg(1-\beta_j^{(3)}-b_j^{(3)}t\bigg)}~.
    \end{split}
    \label{gammasofhfunc}
\end{equation}
The variables $x$ and $y$ are not equal to zero and an empty product is interpreted as unity. The parameters $m_j,n_j,p_j,q_j \, (j=1,2,3)$ are non-negative integers such that $0 \leq m_j \leq p_j$ and $0 \leq n_j \leq q_j$. The Latin letters $a^{(k)}_j,b^{(k)}_j$ are all positive quantities, while the Greek letters $\alpha^{(k)}_j,\beta^{(k)}_j$ are all complex numbers. For a detailed discussion on the convergence of H-functions, we refer the reader to \cite{Hfunc2}.

\paragraph{Generalized I-function} 
The I-function can be regarded as the generalization of the Fox-H functions. They were introduced recently in \cite{Ifunc1,Ifunc2} where their properties have been studied. The I-function  of $r$-variables $\textrm{I}(z_1,\dots,z_r)$ is defined as
\begin{equation}\resizebox{0.9\textwidth}{!}{$%
    \begin{aligned}
        &\textrm{I}^{0,n:\,m_1,n_1;\,\dots;\,m_r,n_r}_{p,q:\,p_1,q_1;\,\dots;\,p_r,q_r}\left(
\begin{array}{c}
z_1 \\ \vdots \\ z_r
\end{array}\middle\vert
\begin{array}{c}
\big(\alpha_j; a_j^{(i)};A_j\big)_{1,p}:\,\big(\gamma_{j}^{(1)}, c_j^{(1)};C_{j}^{(1)}\big)_{1,p_1};\,\dots; \big(\gamma_{j}^{(r)}, c_j^{(r)}; C_{j}^{(r)}\big)_{1,p_r}\\
\big(\beta_j; b_j^{(i)};B_j\big)_{1,q}:\,\big(\delta_{j}^{(1)}, d_j^{(1)};D_{j}^{(1)}\big)_{1,q_1};\,\dots; \big(\delta_{j}^{(r)}, d_j^{(r)}; D_{j}^{(r)}\big)_{1,q_r}\\
\end{array}
\right)  \\
& = \frac{1}{(2 \pi i)^r} \int_{-i \infty}^{+i \infty} \, \dots \,  \int_{-i \infty}^{+i \infty} \, ds_1 \dots ds_r \, \phi(s_1,\dots,s_r) \, \theta_1(s_1) \dots \theta_r(s_r) \, z_1^{s_1} \dots z_r^{s_r} \,~,
    \end{aligned}$}%
    \label{ifunc}
\end{equation}
where $\phi(s_1,\dots,s_r)$, $\theta_i(s_i)$ for $i=1,\dots,r$ are given by
\begin{equation}
    \begin{split}
        \phi(s_1,\dots,s_r) &= \frac{\prod_{j=1}^{n} \Gamma^{A_j} \bigg(1-\alpha_j+\sum_{i=1}^ra_j^{(i)} s_i\bigg)}{\prod_{j=n+1}^{p} \Gamma^{A_j} \bigg(\alpha_j - \sum_{i=1}^{r} a_j^{(i)} s_i\bigg) \prod_{j=1}^q \Gamma^{B_j} \bigg(1-\beta_j+\sum_{i=1}^{r}b_j^{(i)} s_i\bigg)}~,  \\
\theta_i (s_i) &= \frac{\prod_{j=1}^{n_i} \Gamma^{C_j^{(i)}} \bigg(1-\gamma_j^{(i)}+ c_j^{(i)} s_i\bigg) \prod_{j=1}^{m_i} \Gamma^{D_j^{(i)}} \bigg(\delta_j^{(i)} - d_j^{(i)} s_i\bigg)}{\prod_{j=n_i+1}^{p_i} \Gamma^{C_j^{(i)}} \bigg(\gamma_j^{(i)} - c_j^{(i)} s_i\bigg) \prod_{j=m_i+1}^{q_i} \Gamma^{D_j^{(i)}} \bigg(1-\delta_j^{(i)}+ d_j^{(i)} s_i\bigg)}~,
    \end{split}
    \label{gammaifunc}
\end{equation}
where $z_i$ for $i=1,\dots,r$ are not equal to zero and an empty product is interpreted as unity as before. The parameters $n, p, q, m_j,n_j,p_j,q_j \, (j=1,\dots,r)$ are non-negative integers such that $0 \leq n \leq p, q \geq 0, 0 \leq n_j \leq p_j$, and $0 \leq m_j \leq q_j \, (j=1,\dots,r)$ (not all can be zero simultaneously). Crucially for our discussion related to light transforms, the definition of the I-function of $r$-variables presented above holds good even if some of the Latin letters $a_j^{(i)},b_j^{(i)},c_j^{(i)},d_j^{(i)}$ are \textit{zero} or \textit{negative numbers}. In such cases, there exists certain transformation formulae relating them to other I-functions \cite{Ifunc2}. The Greek letters $\alpha_j,\beta_j, \gamma_j^{(i)}, \delta_j^{(i)}$ are all complex numbers and the exponents $A, B, C^{(i)}, D^{(i)}$ of the various Gamma functions appearing in (\ref{gammaifunc}) may assume non-integer values. In all the cases considered in this paper, the exponents will always be \textit{unity}.

\section{Evaluation of \texorpdfstring{${\cal G}(z,\bz)\big|_{\textrm{I}}$}{TEXT}}\label{appen:g1}
In this appendix, we will demonstrate the procedure to evaluate \gi in eq(\ref{equ:G-two-regions}). We again choose to work in the configuration where $0<z<\bar{z}<1$, for which the integral in the first region $\mathcal{G}(z,\bz)\big|_{\textrm{I}}$ further breaks down into three regions:
\begin{equation}\resizebox{0.9\textwidth}{!}{$%
    \begin{aligned}
       &\mathcal{G}(z,\bz)\bigg|_{\textrm{I}} ~=~ \int_{-\infty}^1\,dx\,\frac{1}{x(x-1)}\, {\cal F}_1(z,x)|_{x,z<1}\,{\cal F}_2(x,\bz)|_{x,\bz<1} \\
&= \int_{-\infty}^{z} \frac{dx}{x(x-1)} \, \mathcal{F}_1(z,x)\bigg|_{x<z<1} \, \mathcal{F}_2(x,\bar{z})\bigg|_{x<\bar{z}<1} + \int_{z}^{\bar{z}} \frac{dx}{x(x-1)} \, \mathcal{F}_1(z,x)\bigg|_{z<x<1} \, \mathcal{F}_2(x,\bar{z})\bigg|_{x<\bar{z}<1}  \\
&~~ + \int_{\bar{z}}^{1} \frac{dx}{x(x-1)} \, \mathcal{F}_1(z,x)\bigg|_{z<x<1} \, \mathcal{F}_2(x,\bar{z})\bigg|_{\bar{z}<x<1}~. 
    \end{aligned}$}%
    \label{threeregions}
\end{equation}
Focusing on the second term of (\ref{threeregions}) and using the results in Appendix \ref{appen:four-marked-point-integrals}, we have
\begin{equation}\resizebox{0.9\textwidth}{!}{$%
    \begin{aligned}
        \mathcal{G}(z,\bz)\bigg|_{\textrm{I}} \supset\, & C(\Delta_3,\Delta_4)\, C(\Delta_2,\Delta_4-1)\,|\bar{z}-1|^{\Delta_2+\Delta_4-2} \,\\
&\int_{z}^{\bar{z}} \frac{dx}{x(1-x)} \,|x-1|^{\Delta_2-2} |z-x|^{\Delta_3+\Delta_4-1} |x-\bar{z}|^{\Delta_1-1}  \\
& \times   \, _2F_1 \bigg(2-\Delta_2,\Delta_3,\Delta_3+\Delta_4;-\frac{|x-z|}{|1-x|}\bigg) \, _2F_1 \bigg(1-\Delta_1,\Delta_2,\Delta_2+\Delta_4-1;-\frac{|\bar{z}-1|}{|x-\bar{z}|}\bigg)~.
    \end{aligned}$}%
    \label{thirdtermint2}
\end{equation}
Proceeding as we did in Section \ref{evalutionoffourlighttransforms}, we first recast the Gauss hypergeometric functions as contour integrals over exponents $s_1, s_2$ as in (\ref{mellinbarnesgauss}). We can then perform the $x$-integral as below.
\begin{equation}\resizebox{0.9\textwidth}{!}{$%
    \begin{aligned}
        &\int_{z}^{\bar{z}} \frac{dx}{x(1-x)} \,(1-x)^{\Delta_2-2-s_1} (x-z)^{\Delta_3+\Delta_4-1+s_1} \, (\bar{z}-x)^{\Delta_1-1-s_2}  \\
~=&~ (\bar{z}-z)^{3-\Delta_2+s_1-s_2} \, \bar{z}^{\D_1-s_2-1}z^{s_2-\D_1} \, (1-\bar{z})^{\Delta_2-s_1-3} \, B(\Delta_1-s_2, \Delta_3+\Delta_4+s_1) \\
&~~~~~~~~~~~~~~~\times F_1 \bigg(\Delta_1-s_2,-s_2,3+s_1-\Delta_2,4-\Delta_2+s_1-s_2; -\left|\frac{\bar{z}-z}{\bar{z}}\right|, -\left|\frac{\bar{z}-z}{z(1-\bar{z})}\right|\bigg)
    \end{aligned}$}%
    \label{eq:thirdtermint2xintegral}
\end{equation}
To proceed further, we make use of the following contour integral representation of the Appell $F_1$ hypergeometric function 
\begin{equation}
    \begin{split}
        &F_1(a,b_1,b_2,c;-|z_1|,-|z_2|)  \\ 
&~= \frac{1}{\left(2\pi i\right)^2}\frac{\Gamma(c)}{\Gamma(a) \Gamma(b_1) \Gamma(b_2)}\\
&\qquad \times \int_{-i \infty}^{i \infty}  ds_3 \,ds_4 \, \frac{\Gamma(-s_3) \Gamma(-s_4) \Gamma(b_1+s_3) \Gamma(b_2+s_4) \Gamma(a+s_3+s_4)}{\Gamma(c+s_3+s_4)} \, |z_1|^{s_3} |z_2|^{s_4}~.
    \end{split}
    \label{mellinbarnesofappell}
\end{equation}
The final result is now a four-fold Mellin Barnes integral and can be identified with the $4$-variable I-function 
\begin{equation}\resizebox{0.9\textwidth}{!}{$%
    \begin{aligned}
       \mathcal{G}(z,\bz)\bigg|_{\textrm{I}}  ~&\supset\, |\bar{z}|^{\D_1-1}|z|^{-\D_1}|\bz-z|^{3-\D_2} |1-\bar{z}|^{\Delta_2-\Delta_1-\D_3-1}  \\
& ~\times~\frac{\Gamma(\Delta_3+\Delta_4) \Gamma(\Delta_2+\Delta_4-1)}{\Gamma(2-\Delta_2) \Gamma(\Delta_3) \Gamma(1-\Delta_1) \Gamma(\Delta_2)} C(\Delta_3,\Delta_4) \, C(\Delta_2,\Delta_4-1)  \\ 
&~\times~\textrm{I}^{0,3:1,2;0,2;1,0;1,0}_{3,1:2,2;2,1;0,1;0,1}\left(
\begin{array}{c}
z_1 \\ \vdots \\ z_4
\end{array}\middle\vert
\begin{array}{c}
\big(\alpha_j; a^{(i)}_j;A_j\big)_{1,3}:\,\big(\gamma_{j}^{(1)}, c_j^{(1)};C_{j}^{(1)}\big)_{1,2};\,\dots; \big(\gamma_{j}^{(4)}, c_j^{(4)}; C_{j}^{(4)}\big)_{1,0}\\
\big(\beta_j; b^{(i)}_j;B_j\big)_{1,1}:\,\big(\delta_{j}^{(1)}, d_j^{(1)};D_{j}^{(1)}\big)_{1,2};\,\dots; \big(\delta_{j}^{(4)}, d_j^{(4)}; D_{j}^{(4)}\big)_{1,1}\\
\end{array}
\right)
    \end{aligned}$}%
    \label{ifunctionofint}
\end{equation}
where the entries of the above I-function are:
\begin{itemize}
    \item variables $z_i$: $z_1=\big|\frac{\bar{z}-z}{1-\bar{z}}\big|$, $z_2 = \big|\frac{(1-\bar{z})z}{(\bar{z}-z)\bz}\big|$, $z_3 = \big|\frac{\bar{z}-z}{z}\big|$, and $z_4 = \big|\frac{\bar{z}-z}{z(1-\bar{z})}\big|$~; 
    \item exponents of the Gamma functions: $A_j \,=\,B_j\, =\, C_j^{(i)}\,=\, D_j^{(i)}\, =\, 1$~;
    \item coefficients $c_j^{(i)} = d_j^{(i)}=1$~;
    \item $\a_1=1$, $a_1 = (0,-1,1,0)$~; $\a_2=\D_2-2$, $a_2 = (1,0,0,1)$~;\\
    $\a_3=1-\D_1$, $a_3 = (0,-1,1,1)$~;
    \item $\b_1=\D_2-3$, $b_1=(1,-1,1,1)$~;
    \item $\gamma_{j}^{(1)} = \big(\D_2-1, 1-\D_3 \big) $~; $\gamma_{j}^{(2)} = \big(\D_1, 1-\D_2 \big) $~; 
    \item $\delta_{j}^{(1)} = \big( 0,\D_2-2\big)~$; $\delta_{1}^{(2)} = 2-\D_2-\D_4$~;  $\delta_{j}^{(3)} = \delta_{j}^{(4)} = 0$~.
\end{itemize}
The remaining integrals in (\ref{threeregions}) can be computed in a similar manner and can be shown to be some I-function of \textit{four} variables.

\section{OPEs from the collinear limits of the four-point scalar amplitude}\label{appen:scalar-OPEs}
In this Appendix, we study the various collinear limits of (\ref{equ:scalar-LLOO}).
\paragraph{Scalar - scalar collinear limit} We consider taking the limit $z_{34} \to 0$ starting from a region with $0<z,\zb<1$. We can then make use of (\ref{equ:scalar-LLOO}) and (\ref{equ:F-phi-I}) and the correlator reduces to
\begin{equation}
    \begin{split}
        &\left\langle\,{\bf \bar{L}}[{\cal O}_{\D_1}](z_1,\bar{z}_1){\bf \bar{L}}[{\cal O}_{\D_2}](z_2,\bar{z}_2){\cal O}_{\D_3}(z_3,\bar{z}_3){\cal O}_{\D_4}(z_4,\bar{z}_4)\,\right\rangle \\
        ~\overset{3||4}{\to}&~ N(\b)\,|z_{12}|^{2+\hb-\D_1-\D_2}|z_{14}|^{-2+\D_2-\hb}|z_{24}|^{-2+\D_1-\hb}\\
        &~\Big[1+e^{-\hb i\pi}\Big]\,\Bigg\{\, \frac{C^{(1)}\left(\D_4-\hb,\D_3-\hb\right)}{|\bz_{34}|^{\hb}}\,|\bz_{14}|^{\D_1-2}|\bz_{24}|^{\D_2-2} \\
        &\qquad \qquad ~+~ C(\D_1-1,\D_2-1)\,|\bz_{34}|^{\hb-\D_3-\D_4+1}\,\frac{|\bz_{12}|^{\D_1+\D_2-3}}{|\bz_{14}|^{\D_2-1}|\bz_{24}|^{\D_1-1}}  \,\Bigg\}\\
        &+~ N(\b)\,|z_{12}|^{2-\D_1-\D_2}|z_{14}|^{-2+\D_2}|z_{24}|^{-2+\D_1}|z_{34}|^{-\hb}\\
        &~~[-{\rm sgn}(z_{34}\bz_{34})]^{-\hb}\,\Bigg\{\, \frac{C^{(1)}\left(\D_4-\hb,\D_3-\hb\right)}{|\bz_{34}|^{\hb}}\,|\bz_{14}|^{\D_1-2}|\bz_{24}|^{\D_2-2}\\
        &\qquad \qquad ~+~ C(\D_1-1,\D_2-1)\,|\bz_{34}|^{\hb-\D_3-\D_4+1}\,\frac{|\bz_{12}|^{\D_1+\D_2-3}}{|\bz_{14}|^{\D_2-1}|\bz_{24}|^{\D_1-1}}  \,\Bigg\}~~.\\
    \end{split}
\end{equation}
Note that the leading term is of ${\cal O}\left(z_{34}^0\right)$. This is to be expected as three-point amplitudes involving a massive particle do not produce collinear divergences. In order to read off the OPE, we must first express the parts dependent on coordinates $z_1, z_2, z_3, \zb_1, \zb_2, \zb_3$ as a three-point function. We can do this simply using dimensional analysis. As an example, the fact that
\begin{equation}
    \begin{split}
        &\left\langle\,{\bf \bar{L}}[{\cal O}_{\D_1}](z_1,\bar{z}_1){\bf \bar{L}}[{\cal O}_{\D_2}](z_2,\bar{z}_2)\,{\cal O}_{\D_3+\D_4-\hb,\hb}(z_4,\bar{z}_4) \right\rangle \\
     & \propto |z_{12}|^{2+\hb-\D_1-\D_2}|z_{14}|^{-2+\D_2-\hb}|z_{24}|^{-2+\D_1-\hb}|\bz_{14}|^{\D_1-2}|\bz_{24}|^{\D_2-2}
    \end{split}
    \label{eq:dimanalysis3pt1}
\end{equation}
tells us that the first term in the OPE must contain an operator with conformal dimension $\D_3+\D_4-\frac{\beta}{2}$ and spin $\frac{\beta}{2}$. A similar analysis for the remaining terms lets us write the OPE as $z_{34} \to 0$ in the form
\begin{equation}
    \begin{split}
        {\cal O}_{\D_3}(z_3,\bar{z}_3){\cal O}_{\D_4}&(z_4,\bar{z}_4) ~\sim~  \rho_1\,\frac{C^{(1)}\left(\D_4-\hb,\D_3-\hb\right)}{|\bz_{34}|^{\hb}}{\cal O}_{\D_3+\D_4-\hb,\hb}(z_4,\bar{z}_4) \\
        &\qquad\qquad~+~ \rho_2\,|\bz_{34}|^{\hb-\D_3-\D_4+1}\,{\bf \bar{L}}[{\cal O}_{\D_3+\D_4-\hb,\hb}](z_4,\bar{z}_4) \\
        ~+~&  \rho_3\,[-{\rm sgn}(z_{34}\bz_{34})]^{-\hb}\frac{C^{(1)}\left(\D_4-\hb,\D_3-\hb\right)}{|z_{34}\bz_{34}|^{\hb}}\,{\cal O}_{\D_3+\D_4-\b}(z_4,\bar{z}_4) \\
         &~+~\rho_4\,[-{\rm sgn}(z_{34}\bz_{34})]^{-\hb}\frac{|\bz_{34}|^{\hb-\D_3-\D_4+1}}{|z_{34}|^{\hb}}\,{\bf \bar{L}}[{\cal O}_{\D_3+\D_4-\b}](z_4,\bar{z}_4)~~.
    \end{split}
\end{equation}
$\rho_1$, $\rho_2$, $\rho_3$, and $\rho_4$ can depend on the conformal dimensions but not on $z_3,z_4,\zb_3,\zb_4$.

\paragraph{Scalar light-ray - scalar light-ray collinear limit} Next, we consider the limit $z_{12} \to 0$. Approaching from $0<z<\zb<1$ as above, the correlator reduces to
\begin{equation}
    \begin{split}
        &\langle\,{\bf \bar{L}}[{\cal O}_{\D_1}](z_1,\bar{z}_1){\bf \bar{L}}[{\cal O}_{\D_2}](z_2,\bar{z}_2){\cal O}_{\D_3}(z_3,\bar{z}_3){\cal O}_{\D_4}(z_4,\bar{z}_4)\,\rangle \\
        ~\overset{1||2}{\to}&~ N(\b)\,|z_{12}|^{2+\hb-\D_1-\D_2}|z_{23}|^{-\D_3}|z_{24}|^{-\D_4}\\
        &~\Big[1+e^{-\hb i\pi}\Big]\,\Bigg\{\, \frac{C^{(1)}\left(\D_4-\hb,\D_3-\hb\right)}{|\bz_{34}|^{\hb}}\,|\bz_{23}|^{\hb-\D_3}|\bz_{24}|^{\hb-\D_4}\\
        &\qquad~+~ C(\D_1-1,\D_2-1)\,|\bz_{12}|^{\D_1+\D_2-3}\,\frac{|\bz_{34}|^{\D_1+\D_2-3-\hb}}{|\bz_{23}|^{\hb+1-\D_4}|\bz_{24}|^{\hb+1-\D_3}}  \,\Bigg\}\\
        &~+~ N(\b)\,|z_{12}|^{2-\D_1-\D_2}|z_{23}|^{\hb-\D_3}|z_{24}|^{\hb-\D_4}|z_{34}|^{-\hb}\\
        &~[-{\rm sgn}(z_{34}z_{23}z_{24}\bz_{23}\bz_{24}\bz_{34})]^{-\hb}\,\Bigg\{\, \frac{C^{(1)}\left(\D_4-\hb,\D_3-\hb\right)}{|\bz_{34}|^{\hb}}\,|\bz_{23}|^{\hb-\D_3}|\bz_{24}|^{\hb-\D_4} \\
        &\qquad ~+~ C(\D_1-1,\D_2-1)\,|\bz_{12}|^{\D_1+\D_2-3}\,\frac{|\bz_{34}|^{\D_1+\D_2-3-\hb}}{|\bz_{23}|^{\hb+1-\D_4}|\bz_{24}|^{\hb+1-\D_3}}  \,\Bigg\}~~,\\
    \end{split}
\end{equation}
which implies the OPE
\begin{equation}
    \begin{split}
        {\bf \bar{L}}[{\cal O}_{\D_1}](z_1,\bar{z}_1)&{\bf \bar{L}}[{\cal O}_{\D_2}](z_2,\bar{z}_2)
       ~\sim~ |z_{12}|^{2+\hb-\D_1-\D_2}\Bigg\{\,\rho'_1\,{\bf S}[{\cal O}_{\D_1+\D_2-2-\hb,-\hb}](z_2,\bar{z}_2)\\
       +& \rho'_2\,C(\D_1-1,\D_2-1)\,|\bz_{12}|^{\D_1+\D_2-3}\,{\bf L}[{\cal O}_{\D_1+\D_2-2-\hb,-\hb}](z_2,\bar{z}_2)\Bigg\}\\
        +&|z_{12}|^{2-\D_1-\D_2}\Bigg\{\,\rho'_3\,{\bf S}[{\cal O}_{\D_1+\D_2-2}](z_2,\bar{z}_2) \\
        +&\rho'_4\,C(\D_1-1,\D_2-1)\,|\bz_{12}|^{\D_1+\D_2-3}\,{\bf L}[{\cal O}_{\D_1+\D_2-2}](z_2,\bar{z}_2)\Bigg\}~~.
    \end{split}
\end{equation}

\paragraph{Scalar light-ray - scalar collinear limit} Finally, we consider the limit $z_{13}\to 0$. Since $z \to 1$ in this limit, it is helpful to use hypergeometric identities to rewrite the function ${\cal F}_{\phi,i}(z,\bz)$ in (\ref{equ:F-phi-I}) as
\begin{equation}\resizebox{0.9\textwidth}{!}{$%
    \begin{aligned}
   {\cal F}_{\phi}(z,\bz)|_{z,\bz<1\,\cup\,z,\bz>1} 
         ~=&~  {\cal S}_i(z)\,|z-\bz|^{\hb-\D_3}\,C^{(3)}\left(\D_2-1,\D_3 -\hb \right)\\
         &\quad\times\,_2F_1\left(\D_3-\hb,1-\D_4+\hb,\D_2+\D_3-1-\hb,-\frac{|z-1|}{|\bz-z|}\right)\\
        +&~ {\cal S}_i(z)\,|\bz-z|^{\D_2-2}\,|1-z|^{\D_1+\D_4-2-\frac{\b}{2}}\, C^{(3)}\left(\D_1-1,\D_4 -\hb \right)\\
        &\quad\times\,_2F_1\left(\D_1-1,2-\D_2,\D_1+\D_4-1-\hb,-\frac{|z-1|}{|\bz-z|}\right)
    \end{aligned}$}%
    \label{equ:F-phi-I2}
\end{equation}
which simplifies in the limit $z\to 1$, giving
\begin{equation}\resizebox{0.9\textwidth}{!}{$%
    \begin{aligned}
        &\langle\,{\bf \bar{L}}[{\cal O}_{\D_1}](z_1,\bar{z}_1){\bf \bar{L}}[{\cal O}_{\D_2}](z_2,\bar{z}_2){\cal O}_{\D_3}(z_3,\bar{z}_3){\cal O}_{\D_4}(z_4,\bar{z}_4)\,\rangle ~\overset{2||3}{\to}~ N(\b)\,\Big[1+e^{-\hb i\pi}\Big]\\
        &~\Bigg\{\, \frac{C^{(3)}\left(\D_2-1,\D_3-\hb \right)}{|\bz_{23}|^{\D_3-\hb}}\,|z_{12}|^{\D_4-2-\hb}|z_{24}|^{\D_1-2-\hb}|z_{14}|^{\frac{\D_2+\D_3-\D_1-\D_4}{2}}\,|\bz_{14}|^{\D_1-2}|\bz_{24}|^{2-\D_1-\D_4} \\
        &~+~|z_{23}|^{\D_1+\D_4-2-\hb}|\bz_{23}|^{\D_2-2}\,\frac{C^{(3)}\left(\D_1-1,\D_4-\hb \right)}{|z_{12}|^{\D_1}|z_{24}|^{\D_4}}\,\frac{|\bz_{12}|^{2+\hb-\D_2-\D_3}}{|\bz_{14}|^{\D_4-\hb}|\bz_{24}|^{\hb}}  \,\Bigg\}\\
        &~+~ N(\b)\,[-{\rm sgn}( z_{12}z_{14}z_{24}\bz_{12}\bz_{14}\bz_{24})]^{-\hb}\\
        &~\Bigg\{\, \frac{C^{(3)}\left(\D_2-1,\D_3-\hb \right)}{|\bz_{23}|^{\D_3-\hb}|z_{23}|^{\hb}}\,|z_{12}|^{\D_4-2}|z_{24}|^{\D_1-2}|z_{14}|^{2-\D_1-\D_4}\,|\bz_{14}|^{\D_1-2}|\bz_{24}|^{2-\D_1-\D_4} \\
        &~+~ |z_{23}|^{2-\D_2-\D_3}|\bz_{23}|^{\D_2-2}\,\frac{C^{(3)}\left(\D_1-1,\D_4-\hb \right)}{|z_{12}|^{\D_1-\hb}|z_{14}|^{\hb}|z_{24}|^{\D_4-\hb}}\,\frac{|\bz_{12}|^{2+\hb-\D_2-\D_3}}{|\bz_{14}|^{\D_4-\hb}|\bz_{24}|^{\hb}}  \,\Bigg\}~~.\\
    \end{aligned}$}%
\end{equation}
This gives the corresponding OPE
\begin{equation}
    \begin{split}
        {\bf \bar{L}}[{\cal O}_{\D_2}](z_2,\bar{z}_2){\cal O}_{\D_3}(z_3,\bar{z}_3)&
       ~\sim~ \rho''_1\,\frac{C^{(3)}\left(\D_2-1,\D_3-\hb \right)}{|\bz_{23}|^{\D_3-\hb}}\,{\bf \bar{L}}[{\cal O}_{\D_2+\D_3-\hb,\hb}](z_3,\bar{z}_3)\\
       +& \rho''_2\,|z_{23}|^{\D_1+\D_4-2-\hb}|\bz_{23}|^{\D_2-2}\,{\bf L}[{\cal O}_{\D_2+\D_3-2-\hb,-\hb}](z_3,\bar{z}_3) \\
        +& \rho''_3\,\frac{C^{(3)}\left(\D_2-1,\D_3-\hb \right)}{|\bz_{23}|^{\D_3-\hb}|z_{23}|^{\hb}}\,{\bf \bar{L}}[{\cal O}_{\D_2+\D_3-\b}](z_3,\bar{z}_3)  \\
        +&\rho''_4\, |z_{23}|^{2-\D_2-\D_3}|\bz_{23}|^{\D_2-2}\,{\bf L}[{\cal O}_{\D_2+\D_3-2}](z_3,\bar{z}_3)~~.
    \end{split}
\end{equation}

\section{Conformal block decomposition of the scalar light-ray correlator }\label{appen:CBD-scalar}

In this appendix, we will compute the conformal block decompositions for the term corresponding to the primary operator exchange in the four-point scalar correlator containing two light-ray operators presented in (\ref{equ:scalar-LLOO}). Again, we focus on the region $z,\bz\in[0,1]$. Since (\ref{equ:scalar-LLOO}) already takes the standard four-point function form, the function of cross ratios $G(z,\bz)$ can be read off as
\begin{equation}
    G(z,\bz)|_{z,\bz\in[0,1]} ~=~ \bz^{-\hb}\,(z\bz)^{\frac{\D_3+\D_4}{2}}\,{\cal F}_{\phi}(z,\bz)|_{z,\bz\in[0,1]} ~:=~  {\cal I}_1(z,\bz) ~+~ {\cal I}_2(z,\bz)
\end{equation}
where 
\begin{equation}
    \begin{split}
        {\cal I}_1(z,\bz) ~=&~   S(z)\,\bz^{-\hb}\,(z\bz)^{\frac{\D_3+\D_4}{2}}\,(1-z)^{\hb-\D_3}\, C^{(1)}\left(\D_4-\hb,\D_3-\hb\right)\\
        &\qquad \times \,_2F_1\left(2-\D_2,\D_3-\hb,\D_3+\D_4-\b,\frac{\bz-z}{1-z}\right)\\
        {\cal I}_2(z,\bz) ~=&~ S(z)\,\bz^{-\hb}\,(z\bz)^{\frac{\D_3+\D_4}{2}}\,|\bz-z|^{\D_1+\D_2-3}\,(1-z)^{\D_4-1-\frac{\b}{2}}\,C\left(\D_1-1,\D_2-1\right)\\
        &\qquad\times \,_2F_1\left(\D_1-1,1+\hb-\D_4,\D_1+\D_2-2,\frac{\bz-z}{1-z}\right)
        \end{split}
\end{equation}
and
\begin{equation}
     S(z) ~=~ e^{-\hb i\pi} \,z^{-\hb} ~+~ 1
        ~+~ e^{-\hb i\pi} \, (1-z)^{-\hb}~.
\end{equation}

We focus on the first term, which corresponds to the exchange of quasi-primary operators, and leave the
decomposition of the second term, which corresponds to the exchange of light-ray operators, to future work. We proceed along the lines of \cite{Hu:2022syq,Fan:2021isc} in order to massage ${\cal I}_1(z,\bz)$ into a form from which the conformal block decomposition can be read off. 
\begin{equation}
    \begin{split}
        {\cal I}_1(z,\bz) ~=&~ e^{-\hb i\pi} \,\sum_{n=0}^{\infty}\,\tilde{a}_n\, k^{21}_{34}\left[\,\frac{\D_3+\D_4-\b}{2}+n,\,\frac{\D_3+\D_4-\b}{2}+n\,\right]\\
       ~+&~ \sum_{n,l=0}^{\infty}\, \tilde{b}_{n,l}\,k^{21}_{34}\left[\,\frac{\D_3+\D_4}{2}+n+l,\,\frac{\D_3+\D_4-\b}{2}+n\,\right]\\
       &~+~ e^{-\hb i\pi}\,\sum_{n,l=0}^{\infty}\,\tilde{c}_{n,l} \, k^{21}_{34}\left[\,\frac{\D_3+\D_4}{2}+n+l,\,\frac{\D_3+\D_4-\b}{2}+n\,\right]\\
    \end{split}\label{equ:scalar-CBD}
\end{equation}
where recall that the conformal block is defined as~\cite{Dolan:2011dv}
\begin{equation}
     k^{21}_{34}[h,\bar{h}]~=~z^{h}\,_2F_1(h-h_{12},h+h_{34},2h,z)\,
    \bar{z}^{\bar{h}}\,_2F_1(\bar{h}-\bar{h}_{12},\bar{h}+\bar{h}_{34},2\bar{h},\bar{z})
\end{equation}
and the coefficients are 
\begin{equation}\resizebox{0.9\textwidth}{!}{$%
    \begin{aligned}
        \tilde{a}_n ~=&~ C^{(1)}\left(\D_4-\hb,\D_3-\hb\right)\, \frac{(2-\D_1)_n\,(2-\D_2)_n\,(\D_3-\hb)_n\,(\D_4-\hb)_n}{n!\,(\D_3+\D_4+n-1-\b)_n\,(\D_3+\D_4-\b)_{2n}}\\
       \tilde{b}_{n,l} ~=&~ \tilde{a}_n\,\zeta_l\left(\frac{\D_3+\D_4-\b}{2}+n,\hb\right)\\
       \tilde{c}_{n,l} ~=&~ \tilde{a}_n\,\sum_{k=0}^l\,(-1)^k\,\begin{pmatrix}
       -\hb\\
       k
       \end{pmatrix}\,\zeta_{l-k}\left(\frac{\D_3+\D_4-\b}{2}+n,k+\hb\right)\\
        \zeta_l(h,r)~=&~ \sum_{m=0}^l\frac{(-1)^m}{m!}\frac{(h-h_{12})_{l-m}(h+h_{34})_{l-m}(r+h+l-m-h_{12})_m(r+h+l-m+h_{34})_m}{(l-m)!(2h)_{l-m}(2r+2h+2l-1-m)_m}~.
    \end{aligned}$}%
\end{equation}
We can read off the spectrum of exchanged states in (\ref{equ:scalar-CBD}) to be
\begin{equation}
    \begin{split}
        (h^{(1)},\bh^{(1)}) ~=&~ \left(\,\frac{\D_3+\D_4-\b}{2}+n,\,\frac{\D_3+\D_4-\b}{2}+n\,\right) \\
        (h^{(2)},\bh^{(2)}) ~=&~ \left(\,\frac{\D_3+\D_4}{2}+n+l,\,\frac{\D_3+\D_4-\b}{2}+n\,\right)  \\
    \end{split}
\end{equation}
where $n,l\in \mathbb{Z}_{\ge 0}$.

\bibliography{main2}
\bibliographystyle{JHEP}

\end{document}